\documentclass[9pt,twocolumn,twoside]{opticajnl}
\journal{opticajournal} 

\setboolean{shortarticle}{true}

\usepackage{lineno}

\usepackage{lipsum} 
\graphicspath{{./figs/}}
\title{Tunable 300 W single-frequency 2 micron fiber amplifier}

\author[1,*]{Reagan R. D. Weeks}
\author[1]{Ryan A. Lane}
\author[1]{Brian M. Anderson}

\affil[1]{Air Force Research Laboratory, Directed Energy Directorate, Kirtland AFB, New Mexico 87117, USA}

\affil[*]{reagan.weeks.4@us.af.mil}

\begin{abstract}
We present an all-fiber design for a Tm-based fiber amplifier that can tune over 1992-2065 nm with 300-350 W single-frequency (<100 kHz) output. Over 180 W is achieved out to 2085 nm with <10\% ASE content without utilizing ASE spectral filters. The amplifier employs both Tm- and Tm/Ho-doped gain fibers in two preamplifier stages in addition to longer sections of Tm fiber to extend the bandwidth of the Tm-based high-power amplifier to longer wavelengths (>2050 nm). Efficiencies of $\sim$55\% are realized across the full bandwidth. Roll-off occurs beyond 2085 nm where ASE becomes intractable. The amplifier has an average M$^{2}$ value of 1.39 at high-power due to the presence of light guided within the fiber pedestal. Estimates of the pedestal light and higher-order mode contents are provided.
\end{abstract}

\setboolean{displaycopyright}{false} 

\begin{document}

\maketitle


Thulium-doped fiber amplifiers (TDFAs) have sustained significant interest in the past two decades due to their widespread medical, industrial, and research applications \cite{sincore2017high}. Fiber-based lasers and amplifiers have multiple desirable qualities including high gain and brightness, excellent beam quality and robustness, effective thermal management methods, and relative ability to meet size, weight, and power requirements. TDFAs specifically, with an emission range of 1.9-2.1 \textmu m, offer advantages in atmospheric transmission, nonlinear effects scaling, and improved eye safety over 1 \textmu m systems. \cite{scholle20102}. Their applications include spectroscopy \cite{liao2018dual}, nonlinear generation such as OPO pumping \cite{leindecker2012octave,creeden2008thulium}, LIDAR and remote sensing \cite{li2023development,refaat2016double}, pumping of Ho lasers \cite{budni200350,nvemec2017thulium}, medical procedures \cite{serebryakov2010medical,xie2020brief}, materials processing \cite{mingareev2012welding,mingareev2016principles}, and directed energy \cite{anderson20211}.

Need exists for TDFAs that achieve high power (100s W) with single-frequency linewidths and wavelength tunability out to 2100 nm for studies of atmospheric transmission of high energy lasers at 2 \textmu m. Ideally, such systems would be of an all-fiber design for the advantages of reliability, portability, and ruggedness. TDFAs beyond 2050 nm are desirable due to improved atmospheric transmission, noting however that it is difficult to reach wavelengths beyond $\sim$2050 nm with high output power due to the roll-off of the gain spectrum of Tm$^{3+}$-doped silica \cite{sincore2017high}. Single-frequency plus tunability allow for spectroscopic studies that can probe specific wavelengths to study effects such as thermal blooming. Many TDFAs qualify as narrow-linewidth (typically <$\sim$50 GHz), which is important for applications where coherent beam combination of multiple lasers to achieve higher aggregate powers is needed. The onset of stimulated Brillouin scattering (SBS) is typically the limiting factor for higher power scaling in such systems. Beam combinable systems must balance narrowing the linewidth to increase the coherence length for easier beam combination against broadening the linewidth to raise the power-limiting SBS threshold. For applications such as spectroscopy and coherent sensing (e.g., synthetic aperture LADAR), even narrower linewidths are necessary. Such systems are said to be single-frequency, which we define in the context of TDFAs as having a single longitudinal mode with high side-mode suppression ratio (SMSR) and linewidth less than the spontaneous Brillouin gain bandwidth, which is on the order of 10 MHz for Tm-doped fiber at 2 \textmu m (18 MHz measured in \cite{sincore2017sbs}). Necessarily, the SBS thresholds, and thus attainable power levels, are lower for single-frequency TDFAs.

Previously reported TDFAs that have achieved high output power combined with wide wavelength tunability with narrow linewidths (as opposed to single-frequency) include >100 W over 2030-2050 nm with a 50 GHz linewidth \cite{shah2012integrated}, 90-115 W over 1940-2070 nm with 12-16 GHz linewidth \cite{yin2014high}, >270 W over 1920-2050 nm and >14 GHz linewidth, with 328 W observed at 1930 nm \cite{yin2016300}, and >1 kW tuning over 1943-2050 nm with 40 GHz linewidth \cite{ren2023widely}. The TDFA in Ref. \cite{mccomb2010high} produced 200 W over 1927-2097 nm with <14 GHz linewidth, but this system was not an all-fiber design like the previous references. Examples of high power, single-frequency, all-fiber TDFAs include 102 W at 1971 nm with a <100 kHz source \cite{wang2013102}, 210 W output at 2001 nm with a <2 MHz oscillator \cite{liu2014210}, and 310 W at 1971 nm with <3 MHz \cite{wang2015310}. Goodno et al. achieved 608 W at 2040 nm with <5 MHz linewidth by utilizing a partially free-space design \cite{goodno2009low}. Notably, the all-fiber system developed by Cook et al. generated >80 W across 1920-2010 nm with 114 kHz linewidth, achieving the trifecta of high power, wide tunability, and single-frequency operation \cite{cook2022narrow}. The authors' goal was to design a similar system that operates in the 2.0-2.1 \textmu m region.

\begin{figure*}[htbp]
    \centering\includegraphics[width=\linewidth]{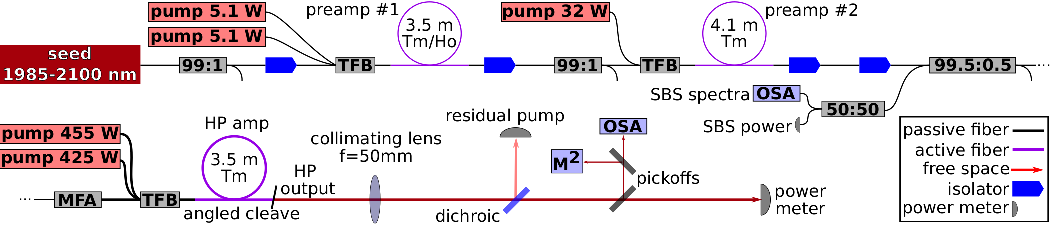}
    \caption{
        Schematic of the TDFA. TFB: tapered-fiber bundle pump/signal combiner; MFA: mode-field adapter; OSA: optical spectrum analyzer; HP: high power.
        }
\label{amplifier_schematic}
\end{figure*}

Here, we present the design and performance of a TDFA in a master oscillator power amplifier (MOPA) configuration which achieves the goals of high power, single-frequency linewidth, and tunability to wavelengths beyond 2050 nm. The all-fiber system operates with <100 kHz linewidth, generating 300-350 W with <1\% amplified spontaneous emission (ASE) content from 1992-2065 nm and >180 W and <10\% ASE out to 2085 nm. All components are commercially available, demonstrating the recent improvements made in readily available 2 \textmu m fiber technologies. The design builds on the techniques used in previous TDFAs and achieves tunability to longer wavelengths by using two novel techniques. First, Tm/Ho co-doped gain fiber is used in the first stage preamplifier. Second, longer sections of Tm-doped fiber are used in the second preamplifier and high power stage to take advantage of ASE reabsorption when operating at longer wavelengths at the expense of lower SBS threshold.

The layout of the MOPA is shown in Fig. \ref{amplifier_schematic}. The seed laser is a narrow-linewidth external cavity diode laser (ECDL, Sacher Lion) tunable from 1985 to 2100 nm, with output powers of 5-10 mW dependent on wavelength. The first stage of preamplification utilizes 3.5 m of co-doped Tm/Ho double-clad fiber (Exail IXF-2CF-TmHo-O-6-130-0.21) with a 6 \textmu m core size and cladding absorption of >3.3 dB/m at 789 nm. The gain fiber is forward-pumped by two pump diodes with a total output of 11.2 W near 793 nm. ASE content constitutes less than 1\% of the output for 2005-2085 nm. The output signal power of the first stage preamplifier ranges from 1.0 to 1.8 W with efficiencies of 10-16\%, with the highest power achieved at 2025 nm. The second-stage preamplifier consists of 4.1 m of Tm-doped fiber (Coherent SM-TDF-10P/130-M, 9 dB/m cladding absorption at 793 nm) forward-pumped by 32 W of pump light near 793 nm. Output signal powers for the second-stage preamplifier range from 16.2 W at 1992 nm to 11.8 W at 2096.75 nm with efficiencies of 37-51\%. The gain fiber and output splice are placed on a cold plate kept at a constant temperature of 18$^{\circ}$ C. A mode-field adapter (MFA) adapts the signal light from the 6 \textmu m core to the 24 \textmu m core of the large-mode area (LMA) fiber used in the high power stage.

The high power amplifier utilizes 3.46 m of Tm-doped LMA double-clad fiber (Coherent, LMA-TDF-25P/250-HE) with a cladding absorption of 11.4 dB/m at 793 nm and featuring a pedestal. The fiber is forward-pumped by two high brightness diodes with a total output of 880 W near 792 nm at full power when operated at 18$^{\circ}$ C. In this configuration, the MOPA is not pump-limited, and the pumps are operated at a lower current resulting in an output power <650 W with a wavelength near 789 nm. The Tm fiber is spooled onto a cold plate with a spiral groove kept at 18 C. The input splice, coated in low index polymer, is placed on the cold plate. The end of the gain fiber is angle-cleaved to $\sim$10$^{\circ}$ to reduce back reflections. So that the entire gain fiber is actively cooled, the fiber output is secured directly onto the cold plate with the final $\sim$2 mm of the fiber extending past the edge of the cold plate. The entirety of the gain fiber is covered in thermal compound to aid in cooling, except for the final $\sim$1 mm with stripped coating before the fiber termination. Neither a passive delivery fiber nor a fiber pump-dump/cladding light stripper are used. Instead, a long-pass dichroic mirror removes and allows measurements of residual pump light from the signal beam in free space directly after a 50 mm focal length 2 \textmu m AR-coated collimating lens. Less than 1 W (representing >99\% pump absorption) of residual pump light was measured at high power across the tunable range.

\begin{figure}[htbp]
    \centering\includegraphics[width=\linewidth]{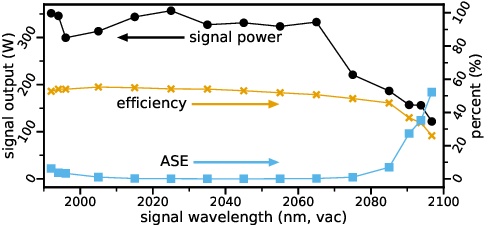}
    \caption{
        Maximum output power of the signal, efficiency, and ASE \% of the MOPA from 1992-2096.75 nm.
        }
\label{output_power_and_efficiency_and_ASE_vs_wavelength}
\end{figure}

Figure \ref{output_power_and_efficiency_and_ASE_vs_wavelength} shows the signal output power of the MOPA across the tunable bandwidth as well as the efficiency at each point based on the pumping power. The output power takes into account the wavelength-dependent losses of the free-space optics before the power meter and the ASE content measured in the total output power. Total gain of the MOPA was nominally 47 dB. The stopping conditions which limited output power varied across the tunable range. For 1992-2045 nm, the backwards Stokes-shifted signal was observed to be 5 dB stronger than the reverse Rayleigh-scattered signal, as measured on an optical spectrum analyzer at the reverse tap before the power amplifier, indicating a strong SBS response. For 2055-2085 nm, the Stokes signal was weaker but the total reverse power was not allowed to exceed 28 mW due to catastrophic failure of the MOPA having occurred at this condition during multiple tests. The MOPA was tested out to 2096.75 nm (due to the presence of two Ho:YAG emission lines at 2090-2097 nm \cite{wang2013resonantly}), where no Stokes peak was observed and further pump increases were pointless since ASE content was >10\% and increasing with pump power.

\begin{figure}[htbp]
    \centering\includegraphics[width=\linewidth]{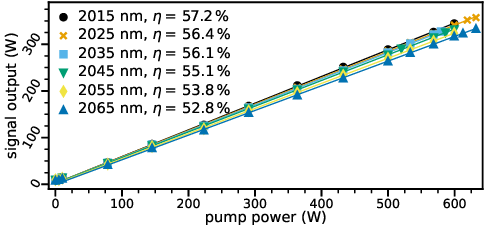}
    \caption{
        Output vs pump power curves for 2015-2065, with y-axis being output power of the signal wavelength. Each is linearly fit to calculate slope efficiency.
        }
\label{output_power_and_slope_efficiency}
\end{figure}

\begin{figure}[htbp]
    \centering\includegraphics[width=\linewidth]{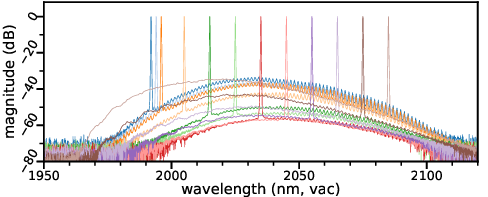}
    \caption{
        Output spectra from 1992-2085 nm at each wavelength's maximum achieved output power. All spectra have been normalized so that peak values are at 0 dB.
        }
\label{output_spectra}
\end{figure}

Over 300 W was achieved from 1992-2065 nm including 351 and 357 W at 1992 and 2025 nm, respectively. Efficiencies, taking into account input signal power, of 46-55\% were measured from 1992-2085 nm, with a strong roll-off occurring at the longest wavelengths due to strong ASE content. ASE content was <1\% for 2015-2065 nm, and <10\% for 1992-2005 and 2075-2085 nm. Slope efficiencies were calculated for 2015-2065 nm, where ASE content was <1\% at all pump powers, and ranged from 53 to 57\% as shown in Fig. \ref{output_power_and_slope_efficiency}. Figure \ref{output_spectra} shows the output spectra for seed wavelengths 1992-2085 nm, with 2090.5-2096.75 nm excluded due to their high ASE content.

High power, single-frequency amplification out to 2085 nm was enabled via two new approaches: using both Tm and Tm/Ho gain fiber in the preamplification stages, and using a longer length of Tm fiber to take advantage of ASE reabsorption. The gain spectrum of Tm/Ho is shifted toward 2100 nm compared to that of Tm, but it is not used exclusively in this design because TDFAs utilizing only co-doped fiber exhibit lower efficiencies (35-40\% observed here) and the inability to tune to shorter wavelengths ($\sim$2000 nm). We desired to have access to narrow atmospheric transmission windows at roughly 1992, 1994, and 1996 nm. In the first preamplifier, Tm-doped fiber was initially tested but produced excessively high amounts of ASE (up to 92\% by power at 2100 nm) when amplifying the long wavelengths of the tunable range. Tm/Ho fiber, in contrast, resulted in 1.5\% ASE content at 2100 nm and 7.1\% at 1985 nm. The same approach was initially used for the second preamplifier. Using Tm/Ho fiber, however, caused the output to be dominated by ASE (up to 97\% by power at 1985 nm) when seeded with wavelengths shorter than 2005 nm. Thus, Tm-doped fiber was used to balance out the gain spectrum, and resulting ASE content, from the first preamplifer. 1.9 m of Tm-doped fiber was tried, but ASE content as high as 23\% was observed at 2100 nm. Using a longer Tm gain fiber length of 4.1 m resulted in lower ASE amounts at longer wavelengths due to reabsorption of the ASE content. At full pump power, 6.3\% and 9.7\% ASE were observed at 1985 and 2100 nm, with 2.4\% and 3.9\% at 1992 and 2097 nm, respectively. The same tactic of using a long length of Tm fiber was used in the high power stage (nearly 40 dB of pump absorption) to again take advantage of ASE reabsorption to improve performance at long wavelengths. A 4.5 m Tm fiber section was first tried, but the onset of SBS at shorter wavelengths limited the output power. The gain fiber was cut back resulting in higher achieved output powers before the onset of SBS at the expense of higher ASE content at the longer wavelengths. The final 3.5 m Tm section balanced these two effects to give the performance reported above.

We note that in an earlier build of the power amplifier, 4.6 m of the same Tm-doped fiber was used with a different pair of pump diodes with a longer center wavelength near 795 nm that were not used in the final build due to being power-limited. At 2090.5 and 2096.75 nm, signal output powers of 177 and 147 W were recorded with lower ASE percentages of 7.22 and 24.9\%, respectively, compared to 157 W/27.3\% and 122 W/52.2\% at comparable pump power for the final build reported above (Fig. \ref{output_power_and_efficiency_and_ASE_vs_wavelength}). No SBS response was observed in either build at these wavelengths. At shorter wavelengths, the SBS response did not become stronger with longer fiber contrary to expectation but was observed to be comparable to the 3.5 m build. We believe that the longer gain fiber promoted reabsorption of the longer wavelength ASE, and that the longer pump wavelength may have reduced the effective gain of the fiber, offsetting the effect of the longer fiber on the SBS response strength. More testing is needed to validate this hypothesis, but these results indicate that improved performance may be achieved with a combination of longer gain fiber and longer wavelength pump diodes. A tunable inline filter would also be expected to lower the amount of ASE seeding the high power stage, likely reducing the ASE content and improving the performance near 2100 nm.

\begin{figure}[htbp]
    \centering\includegraphics[width=\linewidth]{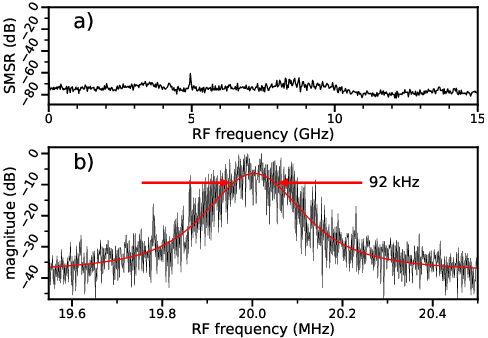}
    \caption{
        (a) Electronic spectrum analyzer trace of the seed at 2040 nm measured by a 10 GHz amplified photodetector. The magnitude of the signal has been calibrated to calculate the SMSR of the seed as a function of frequency spacing from the fundamental longitudinal mode. (b) Lineshape of the seed, as measured via heterodyning with a stabilized frequency comb, with curve-fitting applied to calculate the -3 dB linewidth. 
        }
\label{RFSA_seed}
\end{figure}

Measurements were made to ensure that the seed laser was single-frequency, that is, narrower than the Brillouin gain bandwidth with high SMSR. A trace recorded by an electronic spectrum analyzer shows the RF power spectrum of the seed laser in Fig. \ref{RFSA_seed}(a), which was measured by detecting the seed light with a 10 GHz amplified photodetector. Any secondary lasing modes of the cavity are detected as heterodyne beatnotes since the FSR of the ECDL is <10 GHz. Under most operating conditions, only the noise floor of the detector was measured, indicating excellent single-mode performance. Some wavelength and current pairs did show the presence of a secondary lasing mode, as in Fig. \ref{RFSA_seed}(a) near 5 GHz, but the SMSR was calculated to be 61 dB based on the strength of the heterodyne signal being -66 dBm. This is well below the value at which the MOPA performance would be measurably influenced, such as an increase in the SBS threshold. Figure \ref{RFSA_seed}(b) shows the lineshape of the seed laser's primary lasing mode, which was measured via heterodyning with a stabilized frequency comb (Vescent FFC-100). Since the linewidth of the comb teeth are $\sim$Hz, this measured lineshape is a good reproduction of the seed laser lineshape. A Voigt function was fit to the lineshape and a -3 dB linewidth of 92 kHz was measured. These two metrics indicate the seed laser did qualify as single-frequency.

\begin{figure}[htbp]
    \centering\includegraphics[width=\linewidth]{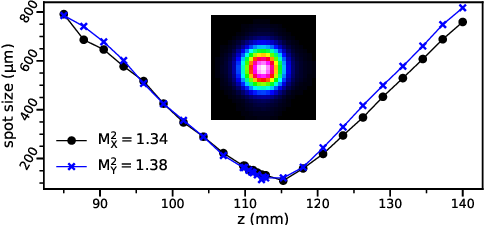}
    \caption{
        M$^{2}$ measurement of high power output seeded at 2025 nm. Inset shows image of beam profile at the focus of the M$^{2}$ measurement with a diameter of 125x121 \textmu m.
        }
\label{M2_2025nm}
\end{figure}

M$^{2}$ measurements were performed at low and high power (Thorlabs M2MS-BP209IR2), and Fig. \ref{M2_2025nm} shows one such measurement at high power for 2025 nm. At high power, M$^{2}$ values of 1.36 were measured at 1992 and 2025 nm and 1.41 at 2065 and 2097.75 nm. A pure fundamental mode output is not achieved likely to due to stray light guided by the pedestal. This is supported by the larger  M$^{2}$ values of 1.49-1.52 measured at low power, which is expected since light in the core will experience more gain compared to light in the pedestal and thus will make up a higher percentage of the output light at high power, improving the value of M$^{2}$. To estimate the amount of light in the pedestal, a pinhole was placed in the collimated free-space beam with a 5.41 mm diameter, which is the expected $\frac{1}{e^{2}}$ diameter of the beam given the wavelength, core radius, core NA, and focal length of the collimating lens. The power transmitted through the iris compared to the total power, measured at multiple wavelengths and preamplifier pumping powers, indicates 10-15\% of the output light is contained within the pedestal. To estimate the higher order mode (HOM) content of the core light, the beam pointing stability of the output was also measured. Averaged across tunable wavelengths, the normalized beam pointing was 2.9\%, increasing to 6.0\% when pressure was applied to the LMA fiber. These values indicate HOM content significantly less than 10\% according to \cite{wielandy2007implications}, and $\sim$1\% or less based on our own mode-solver calculations.

In conclusion, a tunable, single-frequency, all-fiber Tm-based MOPA was developed and tested which produced 300-350 W over 1992-2065 nm, power-limited by the onset of SBS. Performance rolled off at longer wavelengths in terms of output power and ASE percentage, but >180 W was measured out to 2085 nm with <10\% ASE. Efficiency was >49\% over the full bandwidth. Use of Tm/Ho-doped fiber in the first preamplifier in conjunction with longer sections of Tm fiber in the other stages pushed the bandwidth out to 2085 nm. Measurements of the seed laser demonstrated the single-frequency nature of the signal light. M$^{2}$ measurements were also shown with an average value of 1.39 at high power, and estimations of percentage of HOM content and power in the pedestal provided. Future work may utilize similar high power single-frequency TDFAs tunable out to 2100 nm for studies of atmospheric transmission of high energy lasers.

\begin{backmatter}

\bmsection{Funding} This work was funded by the Joint Directed Energy Technology Office (JTO 23-S\&A-0777) and the Air Force Office of Scientific Research (23RDCOR002).

\bmsection{Acknowledgment} We thank Angel Flores, Joel Solomon, and Michael Geraghty for sharing their subject matter and lab expertise.

\bmsection{Disclosures} The authors declare no conflicts of interest. The views expressed are those of the author and do not necessarily reflect the official policy or position of the Department of the Air Force, the Department of Defense, or the U.S. government.

\bmsection{Data Availability Statement} Data underlying the results presented in this paper are not publicly available at this time but may be obtained from the authors upon reasonable request.

\end{backmatter}

\bibliography{2um_SF_amp-OptLett}

\bibliographyfullrefs{2um_SF_amp-OptLett}

\end{document}